\documentstyle[preprint,aps]{revtex}
\begin{document}

\preprint{DOE/ER/40427-6-N96}

\draft

\title{Determination of light-flavor asymmetry in the $\Sigma^{\pm}$
sea by the Drell-Yan process}

\author{Mary Alberg$^{a,b,c}$ and Ernest M. Henley$^{b,c}$}
\address{{\it $^a$ 	Department of Physics,
			Seattle University,
			Seattle, WA 98122, USA}\\
	{\it $^b$ 	Department of Physics,
			Box 351560,
			University of Washington,
			Seattle, WA 98195-1560, USA}\\
	{\it $^c$ 	National Institute for Nuclear Theory,
			Box 351550,\\
			University of Washington,
			Seattle, WA 98195-1550, USA}}

\maketitle

\begin{abstract}
We propose a flavor asymmetry in the light quark sea of
$\Sigma^{\pm}$, which can be measured in Drell-Yan experiments using charged
hyperon beams on proton and deuteron targets. Such a measurement would
help to reinforce the presence of pseudoscalar mesons in a quark model
of baryons.
\end{abstract}


\narrowtext
One of the surprises in the structure of the proton is that the sea
appears to have a flavor asymmetry, an excess of $\bar{d}$ compared to
$\bar{u}$ \cite{NMC91,NMC94,NA51}. Although the experimental results
could also imply an isospin asymmetry, this appears to be less likely,
and we interpret them as an $SU(2)_{Q}$ flavor
asymmetry in the sea \cite{Forte93}. The $\bar{d}$ excess in the
proton is 
expected to be
reflected in an excess of $\bar{u}$ in the neutron; isospin symmetry
would be broken if this were not the case. The evidence for flavor
asymmetry in the proton sea is based on analyses of deep inelastic 
muon scattering \cite{NMC91,NMC94} and Drell-Yan processes \cite{NA51}. One
explanation that has been offered is that the excess of $\bar{d}$
over $\bar{u}$ is due to the Pauli exclusion principle
\cite{Feynman77,Signal8889}. A more likely explanation, in our view,
is that offered by Thomas and colleagues
\cite{Thomas83,Ericson84,Thomas87,Melnitchouk91,Signal91}, Henley and
Miller \cite{Henley90}, and others
\cite{Eichten92,Kumano91a,Kumano91b,%
Hwang91,Szczurek93,Szczurek94,Szczurek96,Holtmann96}, 
namely that the presence of a
pion cloud surrounding a proton favors $\bar{d}$ over $\bar{u}$
because of the excess positive charge of the meson cloud.

It is interesting to apply these arguments to the strange
baryons. Here we focus on the charged $\Sigma^{\pm}$, composed
primarily of $uus$ or $dds$ quarks. We argue that the $\Sigma^{+}$
has an excess of $\bar{d}$ and the $\Sigma^{-}$ has an excess of
$\bar{u}$. The exclusion principle may also play a role
here. Our arguments, however, are based on the pseudoscalar meson cloud
picture. Thus a $\Sigma^{+}(uus)$ will have components
$\Lambda^{0}(uds)\pi^{+}(u\bar{d})$,
$\Sigma^{0}(uds)\pi^{+}(u\bar{d})$,
$\Sigma^{+}(uus)\pi^{0}(\frac{1}{\sqrt{2}}[d\bar{d}-u\bar{u}])$, or
$p(uud)\bar{K^{0}}(\bar{d}s)$; similarly a $\Sigma^{-}(dds)$ can be
$\Lambda^{0}(uds)\pi^{-}(d\bar{u})$,
$\Sigma^{0}(uds)\pi^{-}(d\bar{u})$,
$\Sigma^{-}(dds)\pi^{0}(\frac{1}{\sqrt{2}}[d\bar{d}-u\bar{u}])$, or
$n(udd)K^{-}(\bar{u}s)$. Thus there is a clear favoring of
$\bar{d}$ for $\Sigma^{+}$ and $\bar{u}$ for $\Sigma^{-}$.

There are a number of ways this predicted excess can be
tested. Probably, the most practical is in terms of the Drell-Yan
cross sections for $\Sigma^{\pm}p$ and $\Sigma^{\pm}n$ (i.e. $d$),
e.g., in the inclusive reactions $\Sigma^{\pm}p\rightarrow l^{+} l^{-}
X$, where $l^{\pm}$ are muons or electrons and $X$ is
unmeasured. Beams of $\Sigma^{\pm}$ appear to be adequate for this
purpose. We find that the ratio $\bar{r}_{\Sigma}(x)\equiv
\bar{u}_{\Sigma}(x)/\bar{d}_{\Sigma}(x)$ for the $\Sigma^{+}$ depends on
the known ratios $r_{p}(x)\equiv u_{p}(x)/d_{p}(x)$ and
  $\bar{r}_{p}(x)\equiv \bar{u}_{p}(x)/\bar{d}_{p}(x)$ in the
proton. The former is well-determined from DIS experiments, and the
latter has recently
been determined to be $0.51\pm0.04\pm0.05$ ($\approx 20$ percent
accuracy) at $x\approx 0.18$. We
expect this ratio to be even smaller for the $\Sigma^{+}$ than the proton.

We represent the composition of nucleons and sigma hyperons in terms
of valence and sea quark momentum distributions $q(x)$ and
$\bar{q}(x)$. For clarity, the $Q^{2}$ dependence of these distributions is
suppressed.
We assume isospin reflection symmetry: $u_{p}(x)=d_{n}(x)$,
$\bar{u}_{p}(x)=\bar{d}_{n}(x)$, $u_{\Sigma^{+}}(x)=d_{\Sigma^{-}}(x)$,
$\bar{u}_{\Sigma^{+}}(x)=\bar{d}_{\Sigma^{-}}(x)$, and
$s_{\Sigma^{+}}(x)
=s_{\Sigma^{-}}(x)$.
Drell-Yan cross-sections are proportional to the products
$q(x)\bar{q}(x^{\prime})$, weighted by the product of the quark charges, and
summed over contributions from beam and
target. We neglect sea-quark - sea-quark collisions, which would
contribute below the likely level of accuracy of the experiment. In
the following equations, $q(x)$ represents valence quarks,
$\bar{q}(x)$ represents sea quarks, and the subscript $\Sigma$ refers
to distribution functions for $\Sigma^{+}$. The valence quark
normalizations are: $\int u(x)\,dx = 2$ and $\int d(x)\,dx = 1$.

Consider the Drell-Yan process for $\Sigma^{+}p$. If the experiment is
carried out at $y\approx 0$, then $x_{p}\approx x_{\Sigma}\approx
x$, and

\begin{equation}
\sigma (\Sigma^{+}p)\approx \frac{8\pi\alpha^2}{9\sqrt{\tau}}K(x)
\{\frac{4}{9}[u_{p}(x)\bar{u}_{\Sigma}(x)+u_{\Sigma}(x)\bar{u}_{p}(x)]+
\frac{1}{9}[d_{p}(x)\bar{d}_{\Sigma}(x)+s_{\Sigma}(x)\bar{s}_{p}(x)]\}.
\end{equation}
Then by isospin symmetry
\begin{equation}
\sigma (\Sigma^{-}n)\approx \frac{8\pi\alpha^2}{9\sqrt{\tau}}K(x)
\{\frac{1}{9}[u_{p}(x)\bar{u}_{\Sigma}(x)+u_{\Sigma}(x)\bar{u}_{p}(x)+
s_{\Sigma}(x)\bar{s}_{p}(x)]+\frac{4}{9}d_{p}(x)\bar{d}_{\Sigma}(x)\}.
\end{equation}
We also find
\begin{equation}
\sigma (\Sigma^{+}n)\approx \frac{8\pi\alpha^2}{9\sqrt{\tau}}K(x)
\{\frac{4}{9}[d_{p}(x)\bar{u}_{\Sigma}(x)+u_{\Sigma}(x)\bar{d}_{p}(x)]+
\frac{1}{9}[u_{p}(x)\bar{d}_{\Sigma}(x)+s_{\Sigma}(x)\bar{s}_{p}(x)]\},
\end{equation}
and again by isospin symmetry
\begin{equation}
\sigma (\Sigma^{-}p)\approx \frac{8\pi\alpha^2}{9\sqrt{\tau}}K(x)
\{\frac{1}{9}[d_{p}(x)\bar{u}_{\Sigma}(x)+u_{\Sigma}(x)\bar{d}_{p}(x)+
s_{\Sigma}(x)\bar{s}_{p}(x)]+\frac{4}{9}u_{p}(x)\bar{d}_{\Sigma}(x)\}.
\end{equation}
The factor $K(x)$ accounts for higher-order QCD corrections, and will
factor out in our analysis, since we take ratios of cross
sections. We define a ratio $R(x)$ determined from
the Drell-Yan cross-sections so as to eliminate unknown ratios other
than $\bar{r}_{\Sigma}(x)$ and the recently measured $\bar{r}_{p}(x)$:
\begin{equation}
R(x)\equiv \frac {[\sigma (\Sigma^{+}p)-\sigma(\Sigma^{-}n)]+\bar{r}_{p}(x)
[\sigma (\Sigma^{-}p)-\sigma (\Sigma^{+}n]}
{[\sigma (\Sigma^{+}p)-\sigma (\Sigma^{+}n)]+4[\sigma (\Sigma^{-}p)-\sigma 
(\Sigma^{-}n)]},
\end{equation}
and use Equations 1-4 to write $R(x)$ in terms of the ratios
$\bar{r}_{\Sigma}(x)$, $r_{p}(x)$ and $\bar{r}_{p}(x)$:
\begin{equation}
R(x)=\frac{\bar{r}_{\Sigma}(x)[r_{p}(x)-\bar{r}_{p}(x)]-[1-\bar{r}_{p}(x)
r_{p}(x)]}{5[r_{p}(x)-1]}.
\end{equation}

Thus for $r_{p}(x)\approx 2$ and $\bar{r}_{p}(x)\approx 0.5$,
$R(x)\approx 0.3 \bar{r}_{\Sigma}(x)$.

If $K(x)$ is known, $\bar{d}_{\Sigma}(x)$ can be determined directly
from the cross sections:

\begin{equation}
\bar{d}_{\Sigma}(x)=\frac{27\sqrt{\tau}}{40\pi\alpha^2 K(x)}\frac
{[\sigma (\Sigma^{+}p)-\sigma (\Sigma^{+}n)]+4[\sigma (\Sigma^{-}p)-
\sigma (\Sigma^{-}n)]}{[u_{p}(x)-d_{p}(x)]},
\end{equation}

and $s_{\Sigma}(x)$ can be determined from the cross sections and 
$\bar{s}_{p}(x)$:

\begin{equation}
s_{\Sigma}(x)=\frac{27\sqrt{\tau}}{8\pi\alpha^2 K(x)}\frac
{[\sigma (\Sigma^{+}n)-4\sigma (\Sigma^{-}p)]-r_{p}(x)[\sigma(\Sigma^{+}p)
-4\sigma (\Sigma^{-}n)]}{\bar{s}_{p}(x)[r_{p}(x)-1]}.
\end{equation}

Because of the higher mass of the strange quark, we expect
$s_{\Sigma}(x)$ to peak at a larger $x$ than $d_{p}(x)$.

Quark models with a meson cloud predict the sea quark distributions
$\bar{q}(x)$; they also predict that the difference $D\equiv x
[\bar{d}_{p}(x)-\bar{u}_{p}(x)]$ peaks at $x\approx 0.1$
\cite{Eichten92,Kumano91a,Kumano91b,Hwang91,Szczurek93,Szczurek94,Szczurek96}.
On the basis of meson cloud models\footnote{We are undertaking a calculation
of the $\Sigma^{\pm}$ sea quark distributions.}, the distributions of sea
quarks in
the $\Sigma^{\pm}$ may differ somewhat from those in the nucleon due to
the presence of kaons; this may shift the maximum of $D$ to somewhat
smaller values of $x$. Nevertheless the region of $0\leq x \leq 0.2$
should be a good region for obtaining $\bar{r}_{\Sigma}$.

We believe that the measurement of $R$ should be possible to within $\approx
20\%$ and this is sufficient to establish the preponderance of
$\bar{d}$ over $\bar{u}$ in the $\Sigma^{+}$. As can be seen from
Eq. (6), an error, $e$, in the measurement of $R$ leads to an error of
approximately $3e$ in $\bar{r}_{\Sigma}$. Together with the known
value of $\bar{r}_{p}\approx 0.51$, this measurement would help to
reinforce the validity of the presence of pseudoscalar mesons in a
quark model of baryons, especially if $\bar{r}_{\Sigma}$ is found to
be $\leq 0.5$.

This work has been supported in part by the U.S. Department of Energy,
Contract \# DOE/ER/4027-6-N96, and by the National Institute for Nuclear
Theory. We wish to thank Jen-chieh Peng, Joel Moss, and other
participants in the program INT-96-1, ``Quark and Gluon Structure of
Nucleons and Nuclei'' for helpful discussions.

\end{document}